\documentclass{iau} 
\usepackage{graphicx}

\title[Galactic Centre] 
{The Ecology of the Galactic Centre: \\
 Nuclear Stellar Clusters and  \\
Supermassive Black Holes}

\author[Melvyn B. Davies, Abbas Askar \& Ross P. Church]   
{Melvyn B. Davies$^1$, Abbas Askar$^2$  \and Ross P. Church$^3$}

\affiliation{Lund Observatory, Dept. of Astronomy \& Theoretical Physics \\ Lund University, Box 43,
Lund, SE-221 00, Sweden \\ emails: {\tt $^1$mbd@astro.lu.se}, {\tt $^2$askar@astro.lu.se},
{\tt $^3$ross@astro.lu.se}}

\pubyear{2019}
\volume{351}  
\jname{Star Clusters: From the Milky Way to the Early Universe}
\editors{A. Bragaglia, M.B. Davies, A. Sills \& E. Vesperini, eds.}
\begin{document}

\maketitle

\begin{abstract}
Supermassive black holes are found in most galactic nuclei. A large fraction of these nuclei
also contain a nuclear stellar cluster surrounding the black hole. Here we consider the
idea that the nuclear stellar cluster formed {\it first} and that the supermassive black hole
grew {\it later}. In particular we consider the merger of three stellar clusters to
form a nuclear stellar cluster, where some of these clusters contain a single
intermediate-mass black hole (IMBH). In the cases where multiple clusters contain IMBHs, 
we discuss whether the black holes are likely to merge and whether such mergers
are likely to result in the ejection of the merged black hole from the nuclear stellar cluster.
In some cases, no supermassive black hole will form as any merger product is not
retained. This is a natural pathway to explain those galactic nuclei that
 contain a nuclear stellar cluster but apparently lack a supermassive black hole;
M33 being a nearby example. Alternatively, if an IMBH merger product is retained within the 
nuclear stellar cluster, it may subsequently grow, e.g.\ via the tidal disruption
of stars, to form a supermassive black hole.
\end{abstract}

\firstsection 
\section{Introduction}

Supermassive black holes are found in most, but crucially not all, galactic nuclei.
Perhaps all, and certainly most, nuclei also host a nuclear stellar cluster
at the very centre \cite[(Neumayer \& Walcher 2012, Voggel, et al. 2018, Voggel et al. 2019, Nguyen et al. 2019)]{2012AdAst2012E..15N,2018ApJ...858...20V,2019ApJ...871..159V,2019ApJ...872..104N}.  Intriguingly, some galactic nuclei are found to contain a nuclear
stellar cluster but show no evidence for a supermassive black hole. M33 is a local
example. This galaxy is bulge-free, which suggests  a relatively quiet history without any major mergers.
We therefore  conclude that M33 likely has {\it never} possessed a supermassive black hole. 
Could it be therefore that nuclear clusters form first, and supermassive black holes then 
grow in a subset of these clusters?

\begin{figure}[t]
\begin{center}
\includegraphics[width=3.0in]{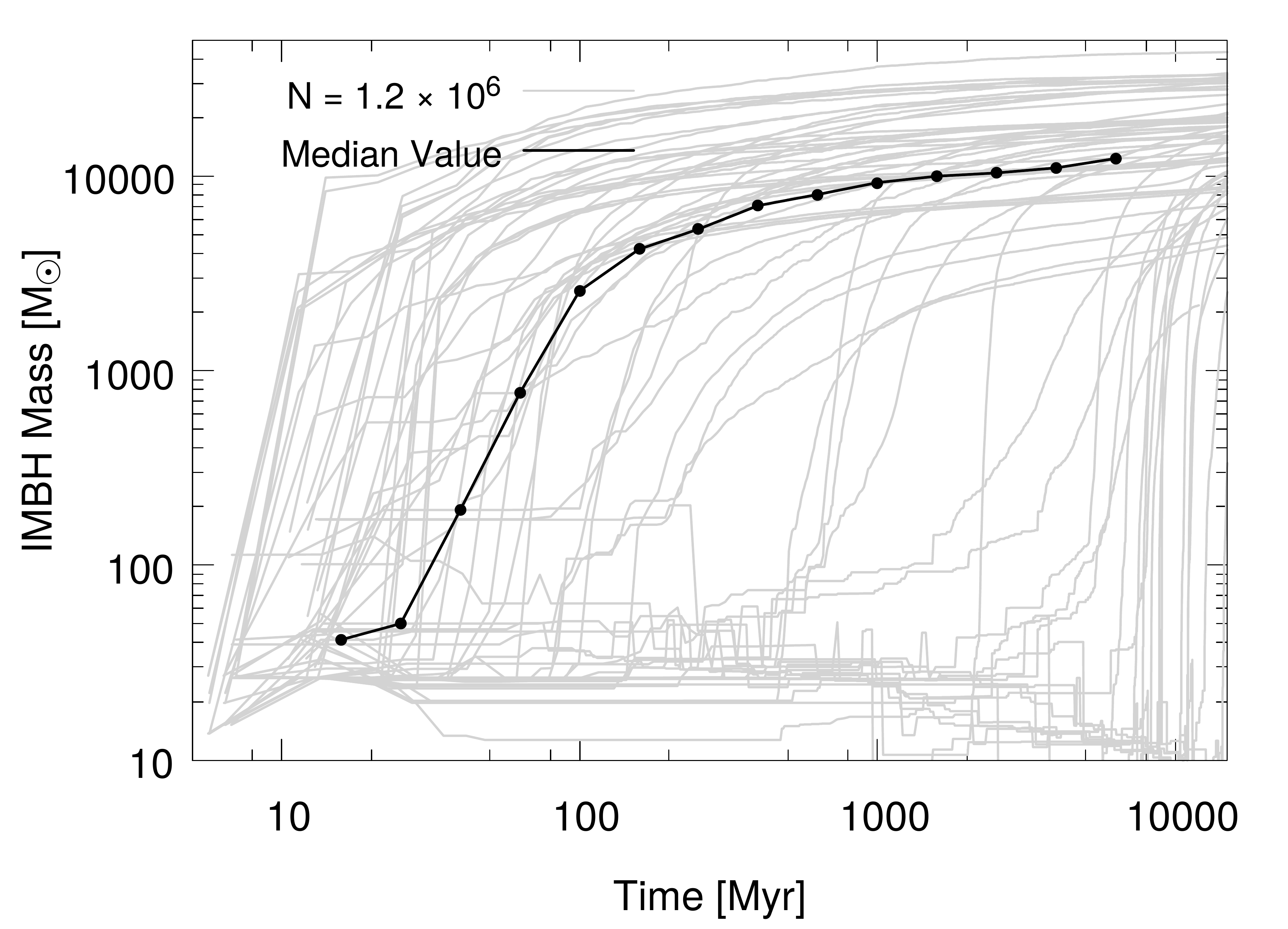} 
 \caption{Mass of intermediate mass black holes as a function of time produced
 in clusters initially containing $1.2 \times 10^6$ stars as determined using
 the Monte-Carlo N-body code {\it MOCCA}.}
   \label{mbdavies_figure1}
\end{center}
\end{figure}

Nuclear stellar clusters may form via the inspiral and merger of smaller stellar
clusters \cite[(Antonini et al. 2012, Mastrobuono-Battisti et al. 2014)]{2012ApJ...750..111A,2014ApJ...796...40M}. 
If a stellar cluster is sufficiently dense, an intermediate-mass black hole (IMBHs, having
a mass between 100 and 1000 M$_\odot$) may form via either 
runaway collisions or via the growth of a stellar-mass black hole through
the tidal disruption of stars \cite[(Portegies Zwart et al. 2004, Freitag et al. 2006, Stone et al. 2017)]{2004Natur.428..724P,2006MNRAS.368..141F,2017MNRAS.467.4180S}. The outcome 
of Monte-Carlo $N$-body simulations of clusters initially containing $1.2 \times 10^6$ objects 
are shown in Fig.\,\ref{mbdavies_figure1}. In considering a broad range
of initial cluster properties, about 25 per cent produce IMBHs \cite[(Askar et al. 2017, Arca Sedda et al. 2019)]{2017MNRAS.464.3090A,2019arXiv190500902A}.

To investigate how often supermassive black holes 
are likely to form and grow within nuclear stellar clusters, 
we have performed a series of numerical simulations 
of the merger of three stellar clusters.
We consider a range of cases where various combinations of the three clusters
contain IMBHs. As shown below, the outcomes 
will depend on the number of IMBHs contained within all three clusters.

\begin{table}[b]
  \begin{center}
  \caption{Table showing the simulated merging stellar cluster models. Each simulated model had three merging star clusters, a central cluster (C) comprising 50000 stars, an infalling cluster (IF1) comprising 30000 stars and a second infalling cluster (IF2) comprising 15000 stars. Each of these star clusters were Plummer models with a half-mass radius of about 2.4 pc. The table lists the simulated models, number and mass of the IMBHs they contain, the initial position of the IMBH(s) and the time up to which the models were simulated.}
  \scriptsize{
  \begin{tabular}{|l|c|c|c|c|}\hline 
{\bf Models} & {\bf IMBH Number and Mass} & {\bf Total Mass of} & {\bf Initial Location} & {\bf Evolution Time} \\ 
   &  & {\bf Merging Clusters [$M_{\odot}$]} & {\bf of IMBH(s)} &  {\bf [Myr]} \\
 0.1 & No IMBH & $\rm 8.4 \times 10^{4}$ & None & 375 \\ \hline 
1.1 & 1 - 1000 $M_{\odot}$ & $\rm 8.5 \times 10^{4}$ & C & 101 \\ \hline
2.1 & 2 - 1000 $M_{\odot}$ and 500 $M_{\odot}$ & $\rm 8.55 \times 10^{4}$ & C and IF1 & 135.3 \\ \hline
2.2 & 2 - 1000 $M_{\odot}$ and 100 $M_{\odot}$ & $\rm 8.5 \times 10^{4}$ & C and IF1 & 75 \\ \hline
2.3 & 2 - 1000 $M_{\odot}$ and 200 $M_{\odot}$ & $\rm 8.52 \times 10^{4}$ & C and IF1 & 114.3 \\ \hline
2.4 & 2 - 500 $M_{\odot}$ and 200 $M_{\odot}$ & $\rm 8.47 \times 10^{4}$ & IF1 and IF2 & 173.5 \\ \hline
3.1 & \begin{tabular}[c]{@{}c@{}}3 - 1000 $M_{\odot}$, 500 $M_{\odot}$,\\  200 $M_{\odot}$\end{tabular} & $\rm 8.57 \times 10^{4}$ & C, IF1 and IF2 & 195 \\ \hline
3.2 & 3 - 500 $M_{\odot}$, 1000 $M_{\odot}$, 200 $M_{\odot}$ & $\rm 8.57 \times 10^{4}$ & C, IF1 and IF2 & 66.3 \\ \hline
  \end{tabular}
  }
 \end{center}
    \label{mbdavies_table1}
\end{table}

\begin{figure}[h]
\begin{center}
\includegraphics[width=5.0in]{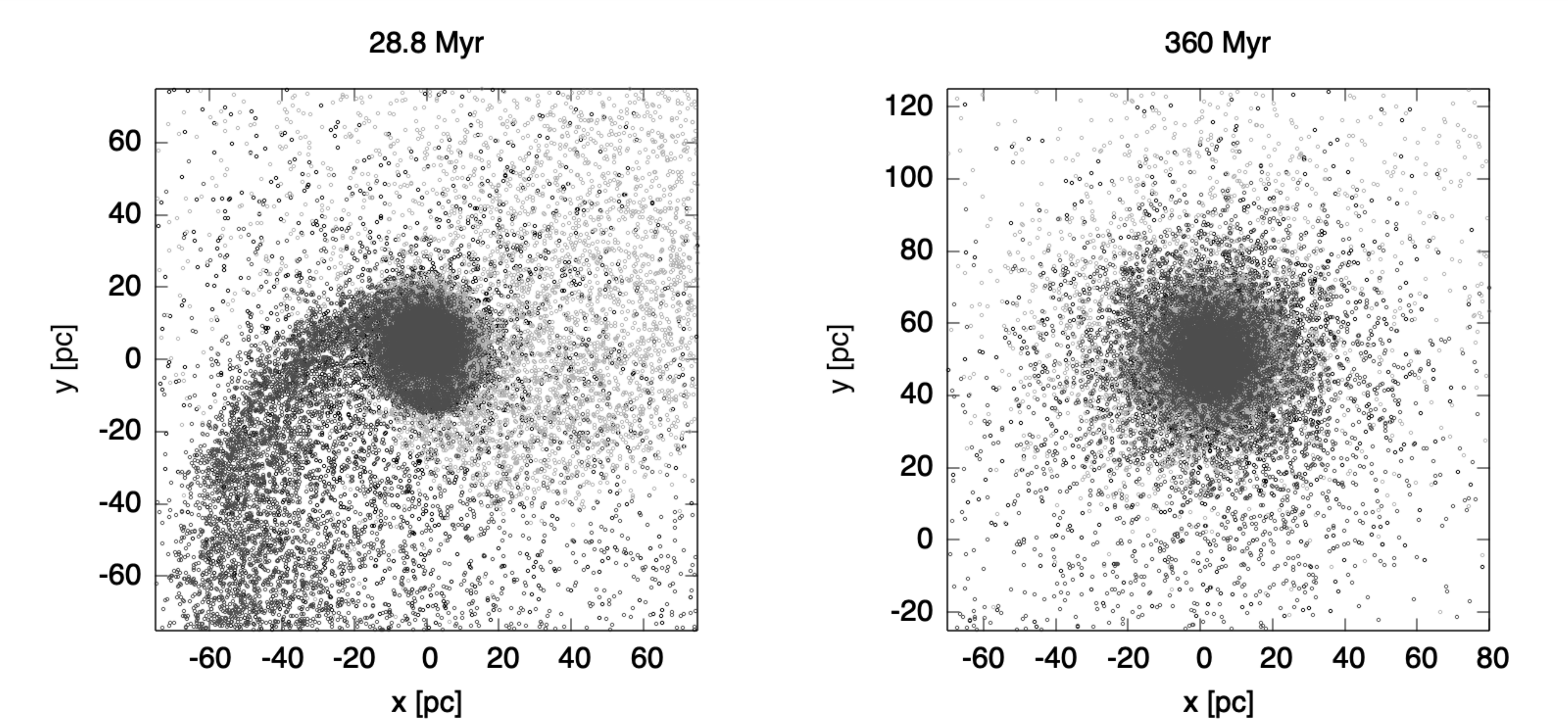} 
 \caption{Particle plot showing the evolution of thee stellar clusters spiralling together
 to form a single nuclear stellar cluster (model 0.1 -- containing zero intermediate-mass black holes).}
   \label{mbdavies_figure2}
\end{center}
\end{figure}

\section{Possible Outcomes of Cluster Mergers}

Our runs are listed in Table \ref{mbdavies_table1}. In all simulations we modelled 
the merger of three stellar clusters to form a nuclear stellar cluster.
We considered simulations containing zero, one, two and three intermediate-mass black holes
(IMBHs). Possible outcomes are described below as a function of the number 
of intermediate-mass black holes (IMBHs) contained within all three clusters. 
 
\begin{figure}[b]
\begin{center}
\includegraphics[width=3.0in]{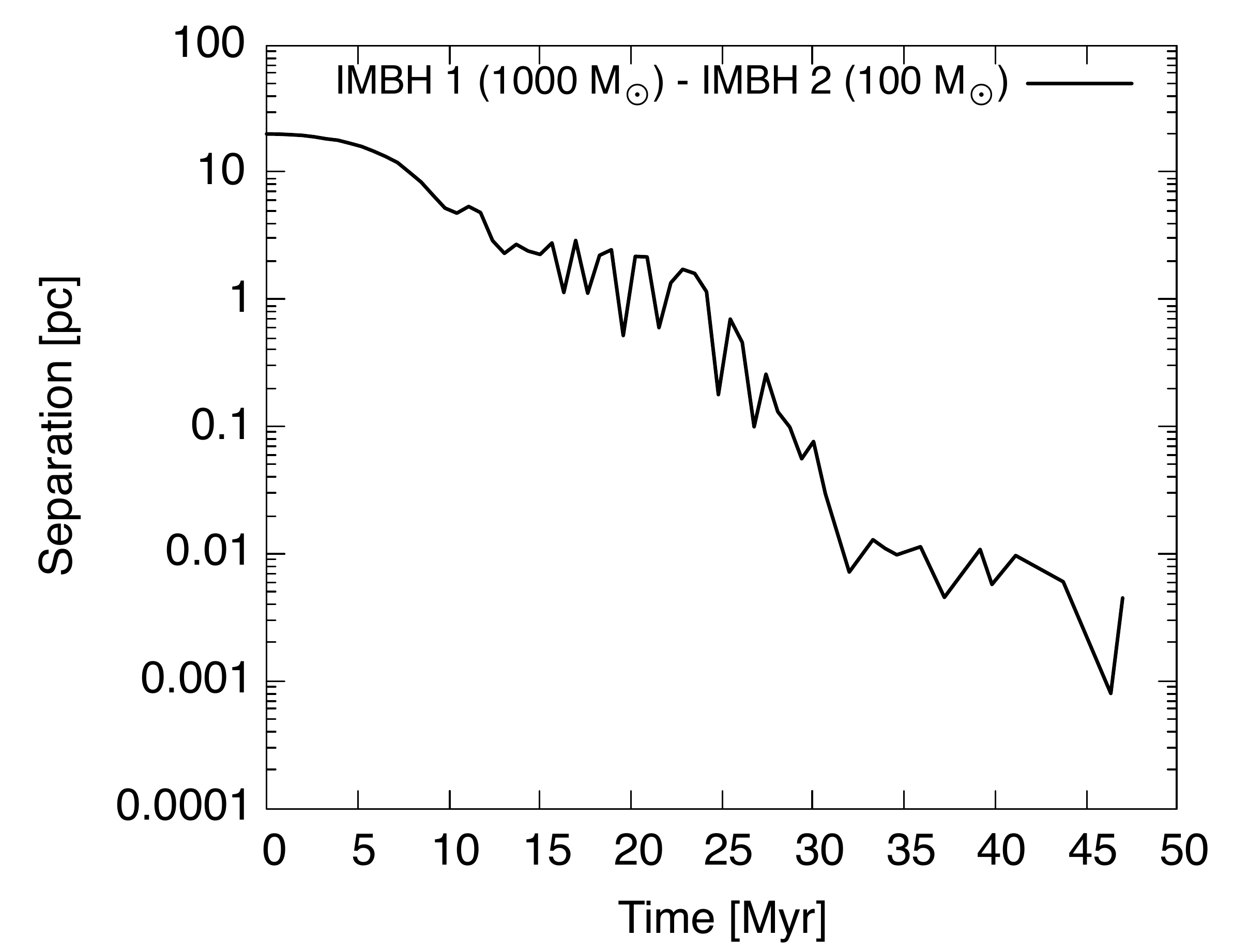} 
 \caption{The instantaneous separation of the two IMBHs as a function of time
 for Model 2.2. After 30 Myr, the two black holes form a tight binary with essentially
 no  stars  between them. The non-smooth structure  is due to the orbital eccentricity.}
   \label{mbdavies_figure3}
\end{center}
\end{figure}
{\underline{\it Zero IMBHs}}. In Fig.\,\ref{mbdavies_figure2}, we show particle plots of 
model 0.1 where we model the merger of three clusters where no clusters contain an IMBH.  
The clusters are merge quickly (in only a few orbits) producing a a cylindrically
symmetric nuclear stellar cluster containing roughly 85 per cent of the original stars.

{\underline{\it One IMBH}}. In model 1.1, we consider the case where the central cluster
contains a 1000 M$_\odot$ black hole. The merger follows in a very similar way to that
seen in model 0.1. In other words, the IMBH is not found to have an immediate impact
on the cluster. In at least some cases, the IMBH will later grow via 
stellar tidal disruptions.

{\underline{\it Two IMBHs}}. We consider four combinations of IMBH masses and initial locations.
 In all cases, we find that the separation of
the two IMBHs decreases significantly over time as they make their way
into the central regions of the nuclear stellar cluster. They form a distinct binary after
about 30 Myr. Such binaries will likely merge via the emission of gravitational radiation
and the scattering of stars. A recoil kick is likely to eject the merged black hole from 
the nuclear stellar cluster, unless the secondary mass is less than about 10 per cent of the 
primary \cite[(Baker et al. 2008, Morawski et al. 2018)]{2008PhRvD..78d4046B,2018MNRAS.481.2168M}.

{\underline{\it Three IMBHs}}. The results for these runs are closely related
to what is seen above for two IMBHs. Two IMBHs typically form a binary first whilst
the third black hole is left on a wider orbit. If the outer binary is ground down, 
the system may form a distinct triple (i.e. an object devoid of stars in the space
between the IMBHs). The Kozai mechanism may then accelerate the merger
process as the inner binary is driven to high eccentricities \cite[(Miller \& Hamilton 2002)]{2002ApJ...576..894M}.
Alternatively, binary-single encounters involving all three IMBHs may lead
to the ejection of one or all three IMBHs \cite[(Sigurdsson \& Phinney 1993, Sigurdsson \& Hernquist 1993, Kulkarni et al. 1993, Davies et al. 1994)]{1993ApJ...415..631S,1993Natur.364..423S,1993Natur.364..421K,1994ApJ...424..870D}.


\section{Summary}

{\underline{\it SMBHs may form after NSCs}}. 
We have shown how a nuclear stellar cluster (NSC) may form first
via the merger of stellar clusters which then go on to form and
grow a supermassive black hole (SMBH).

{\underline{\it IMBHs may form and grow in stellar clusters}}. 
Intermediate-mass black holes may form and grow within stellar clusters
before the clusters themselves merge to form a nuclear stellar cluster.

{\underline{\it 1 + 1 = 0 }}. Merging black-hole binaries may be ejected from nuclear stellar clusters
as they receive a recoil kick due to the asymmetric emission of gravitational radiation.

{\underline{\it 2 + 1 = 2 or 0}}. Encounters between an IMBH binary and 
a single IMBH may result in the ejection of one, or all three, IMBHs.

{\underline{\it 1 + small = larger}}. When two IMBHs merge, 
if the secondary mass  is less than about 10 per cent of the primary,
the recoil kick will be sufficiently small, that the merger product will be retained
by the NSC. It may then grow into a supermassive black hole.

\section{Acknowledgements}

We would like to thank Long Wang for his assistance in using NBODY6++GPU that was used to simulate the merging clusters. AA is supported by the Carl Tryggers Foundation for Scientific
Research through the grant CTS 17:113. RC is supported by grant 2017-04217 from the Swedish Research Council.

\end{document}